\documentclass[aps,prb,superscriptaddress,twocolumn]{revtex4-2}

\usepackage{graphics}
\usepackage{graphicx}% Include figure files
\usepackage{dcolumn}% Align table columns on decimal point
\usepackage{bm}% bold math
\usepackage[usenames]{color}

\graphicspath{{./figs/}}
\def\TN{$T_\mathrm{N}$}
\def\Rw{$R_w$}

\def\muB{$\mu_{\mathrm{B}}$}

\def\MPSe{MnPSe$_{3}$}
\def\MPS{MnPS$_{3}$}
\def\Se2{MnPSSe$_{2}$}

\def\S2{MnPS$_{2}$Se}
\def\An{$\mathrm{\AA}$}

\begin{document}

\title{	Local spin structure in the layered van der Waals materials MnPS$_{x}$Se$_{3-x}$}

	\author{Raju Baral}
        \email{baralr@ornl.gov;lead contact}
	\affiliation{ %
		Neutron Scattering Division, Oak Ridge National Laboratory, Oak Ridge, Tennessee 37831, USA.
	} %
	
\author{Amanda~V.~Haglund}
\affiliation{Department of Materials Science and Engineering, University of Tennessee, Knoxville, TN 37996, USA.}

\author{Jue~Liu}
	\affiliation{ %
		Neutron Scattering Division, Oak Ridge National Laboratory, Oak Ridge, Tennessee 37831, USA.
	} %

\author{Alexander~I.~Kolesnikov}
	\affiliation{ %
		Neutron Scattering Division, Oak Ridge National Laboratory, Oak Ridge, Tennessee 37831, USA.
	} %

%	\author{Benjamin A. Frandsen}
%	\affiliation{ %
%		Department of Physics and Astronomy, Brigham Young University, Provo, Utah 84602, USA.
%	} %

\author{David~Mandrus}
\affiliation{Department of Materials Science and Engineering, University of Tennessee, Knoxville, TN 37996, USA.}
\affiliation{Materials Science and Technology Division, Oak Ridge National Laboratory, Oak Ridge, TN 37831, USA.}

	\author{Stuart Calder}
        \email{caldersa@ornl.gov}
	\affiliation{ %
		Neutron Scattering Division, Oak Ridge National Laboratory, Oak Ridge, Tennessee 37831, USA.
	} %

\begin{abstract}
 Two-dimensional (2D) layered materials, whether in bulk form or reduced to just a single layer, have potential applications in spintronics and  capacity for advanced quantum phenomena. A prerequisite for harnessing these opportunities lies in gaining a comprehensive understanding of the spin behavior in 2D materials. The low dimensionality motivates an understanding of the spin correlations over a wide length scale, from local to long range order.  In this context, we focus on the magnetism in bulk \MPSe ~and \MPS,  2D layered van der Waals antiferromagnetic semiconductors. These materials have similar honeycomb Mn layers and magnetic ordering temperatures, but distinct spin orientations and exchange interactions. We utilize neutron scattering to gain deeper insights into the local magnetic structures and spin correlations in the paramagnetic and ordered phases by systematically investigating a MnPS$_{x}$Se$_{3-x}$ ($x$ = 0, 1, 1.5, 2, 3) series of powder samples using total neutron scattering measurements. By employing magnetic pair distribution function (mPDF) analysis, we unraveled the short-range magnetic correlations in these materials and explored how the non-magnetic anion S/Se mixing impacts the magnetic correlations. The results reveal that the magnetism can be gradually tuned through alteration of the non-magnetic S/Se content, which tunes the atomic structure. The change in magnetic structure is also accompanied by a control of the magnetic correlation length within the 2D honeycomb layers. Complimentary inelastic neutron scattering measurements allowed a quantification of the change in the magnetic exchange interactions for the series and further highlighted the gradual evolution of spin interactions in the series MnPS$_{x}$Se$_{3-x}$.
\end{abstract}
	
\maketitle

\section{Introduction}

%%%%%%%%%%%%%%%%%
% Begin Figure
%%%%%%%%%%%%%%%%%
\begin{figure*}
        \centering
	\includegraphics[width=180mm]{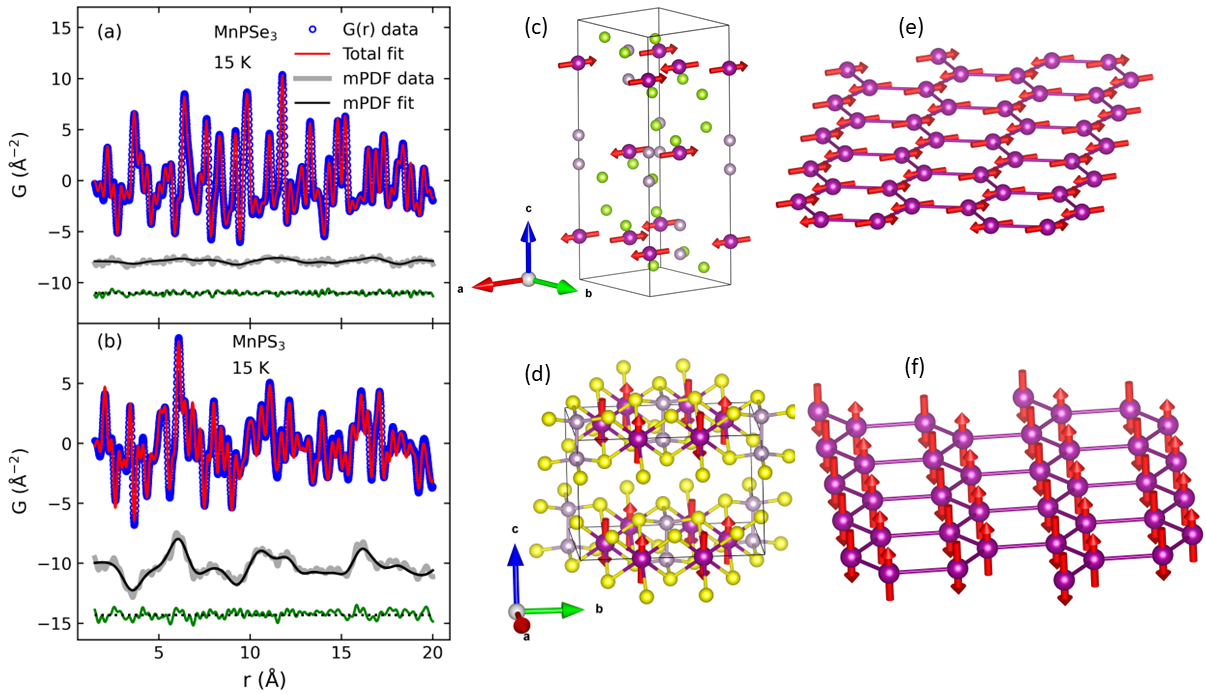}
	\caption{\label{fig:mPDF_pure_15K} The total neutron PDF (including both magnetic and nuclear scattering) and isolated magnetic PDF of (a) \MPSe~and (b) \MPS~at 15~K. The blue symbols and red curve represent the experimental total PDF data and total PDF fit, respectively. The gray and black curves represent the isolated magnetic PDF data and magnetic PDF fit (vertically offset for clarity). The green curve represents the overall fit residual.  (c) Crystal structure of \MPSe~ with Mn spins pointing in $ab$-plane (space group=R$\bar{3}$). (d) Crystal structure of \MPS~ with Mn spins pointing out of the $ab$-plane (space group=C${2}/m$). The hexagonal motif of the Mn ions for (e) \MPSe~ and (f) \MPS~. Purple, gray, green, and yellow spheres represent the Mn, P, Se, and S atoms respectively.
	}		
\end{figure*}
%%%%%%%%%%%%`
% End Figure
%%%%%%%%%%%%

Reducing the dimensionality of a compound down to topologically constrained layers can uncover new and intriguing behaviors that go beyond well-established classical understanding. A prime example is graphene, in which a single layer is extracted from the weakly connected van der Waals (VDW) bonded layers in graphite \cite{geim2013van}. Graphene and other 2D materials have undergone significant interest stemming from their potential to exhibit novel quantum effects, such as Dirac semi-metal and quantum anomalous Hall insulating behavior \cite{zhang2005experimental,yu2010quantized}. Interesting properties can be found in materials ranging from the ideal two-dimensional (2D) monolayers to qausi-2D materials with more complex bulk VDW-bonded structures \cite{Burch2018}. Developing functional spintronic materials typically relies on harnessing magnetic and semiconducting properties, however, graphene lacks these phenomena. As a result, 2D VDW materials that have intrinsic magnetism and semiconductivity with a honeycomb arrangement are of interest. A prerequisite for developing functional magnetic materials and uncovering novel quantum behavior in 2D materials lies in gaining a comprehensive understanding of the spin behavior over the relevant length scales.

\MPS~ and \MPSe~ are layered VDW antiferromagnetic materials that posses both magnetic and semiconducting properties, with the magnetic ions adopting the same hexagonal motif as graphene. Although each crystallizes in different space groups,  \MPS~ in C$2/m$ and \MPSe~ in R$\bar{3}$, they have nearly identical N\'{e}el temperatures of 78 K for \MPS~ and 74 K for \MPSe~\cite{takano2004magnetic, wiedenmann1981neutron}. Both of these compositions contain an equivalent fraction of magnetic ions ($n_{s}$ = 1/5) within their unit cells and are S = 5/2 materials. Despite their similar transition temperatures and magnetic ion content, intriguing distinctions become apparent in their magnetic behavior. \MPS~ has Heisenberg antiferromagnetic spins that point out of the hexagonal $ab$-plane \cite{ressouche2010magnetoelectric}, while \MPSe~ displays XY anisotropy in its magnetic behavior with spins lying in the $ab$-plane \cite{calder2021magnetic}. The crystal structures with spin directions are shown in Fig.\ref{fig:mPDF_pure_15K}.

By altering the non-magnetic S and Se ion content it is possible to chemically tune between \MPS~ and \MPSe~ in the series MnPS$_{x}$Se$_{3-x}$. This provides a way to explore and understand the evolution between the two magnetic states. Stabilization of this series has been shown for a wide concentration of $x$ values for powders and single crystal samples \cite{han2023emergent,hillman2011structural,hou2020alloy,yan2011synthesis}. Based on bulk characterizations, these studies have found indications of a change in the magnetic ordering within the series as $x$ is altered. This motivates further investigation to probe the microscopic spin states and magnetic exchange interactions. 

In this work, we present total neutron scattering and inelastic neutron scattering measurements to achieve a comprehensive picture of the long-range and short-range correlations and exchange interactions in the series MnPS$_{x}$Se$_{3-x}$ (x = 0, 1, 1.5, 2, 3). Using magnetic pair distribution function (mPDF) analysis of total neutron scattering data, we present the real-space visualization of local spin-spin correlations in the paramagnetic state. The neutron scattering data reveals signatures of short-range magnetic correlations well above the N\'{e}el temperature for \MPS~and we modeled this with mPDF analysis to extract the short- and long-range spin states. As the S content increases towards \MPSe~the quasi-2D behavior is reduced, as shown by the decrease in the magnetic correlation length. This behavior is confirmed by inelastic neutron scattering measurements that show an increase in the interlayer exchange interactions as the Se content increases. We show that control of the non-magnetic S and Se content provides a sensitive mechanism to tune the magnetic spin direction between in-plane and out-of-plane through alteration of the crystal structure and small modifications in the Mn-Mn distances. This suggests routes beyond chemical pressure, such as strain, could provide promising routes to tune the magnetism in layered materials. Moreover the methodology employed of mPDF, that can access various spin length scales over a wide temperature range, is shown to be well suited to 2D VDW materials that can intrinsically contain competing dimensionality of spin ordering.

\section{Experimental methods}
\subsection{Sample synthesis}

Polycrystalline samples of \MPSe ~and \MPS~ were synthesized following the standard solid-state reaction method as described in Ref.~\cite{calder2021magnetic}. The samples MnPSSe$_{2}$,~MnPS$_{1.5}$Se$_{1.5}$,~and MnPS$_{2}$Se were similarly synthesized using the identical method with suitable precursor materials. Confirmation of sample quality was accomplished through laboratory X-ray diffraction analysis.

\subsection{Neutron total scattering}

Neutron total scattering experiments were conducted on both the Nano Scale-Ordered Material Diffractometer (NOMAD) and the HB-2A powder diffractometer. NOMAD is a dedicated total scattering instrument for PDF analysis. The HB-2A powder diffraction instrument is traditionally utilized for reciprocal space analysis at low $Q$, however here we use a short wavelength to access higher $Q$ and perform total scattering measurements and PDF analysis. With the capabilities to measure samples at millikelvin temperatures and apply magnetic fields HB-2A can provide unique conditions for magnetic PDF analysis for a broad range of quantum material research.

For both the HB-2A and NOMAD data the atomic PDF fits were performed using PDFgui~\cite{farro;jpcm07} and the magnetic PDF fits were obtained using diffpy.mpdf~\cite{Frandsen:diffpy.mpdf}, a Python package designed for magnetic PDF calculations and fitting.

\subsubsection{NOMAD}

Measurements on \MPSe~and \MPS~were conducted at NOMAD, located at the Spallation Neutron Source (SNS), Oak Ridge National Laboratory (ORNL)~\cite{neuef;nimb12}. These polycrystalline samples (\MPSe ~and \MPS) were loaded into 6 mm vanadium cans in a helium glove box. Total scattering patterns were collected in a helium cryostat over a range of temperatures from 15 K to 300 K. The collected data was reduced and Fourier transformed with a \textit{Q}$\rm_{max}$ value of 20 $\rm \AA$$^{-1}$, employing automatic data reduction scripts at NOMAD  \cite{mcdonnell2017addie}. The choice of \textit{Q}$\rm_{max}$ value is critical. A high \textit{Q}$\rm_{max}$ value can significantly improve the resolution in real space, but it also introduces statistical noise and artifacts. Consequently, a \textit{Q}$\rm_{max}$ value of 20 $\rm \AA$$^{-1}$ was strategically selected to optimize the balance between achieving high resolution and minimizing noise artifacts.

\subsubsection{HB-2A}

Neutron total scattering patterns of the doped samples (MnPSSe$_{2}$, MnPS$_{1.5}$Se$_{1.5}$, and MnPS$_{2}$Se) were collected on the HB-2A powder diffractometer at the High Flux Isotope Reactor (HFIR) at ORNL\cite{garlea2010high}. A vertically focusing germanium monochromator was used to select the wavelength of 1.12~\AA~from the Ge(117) reflection. Approximately 5 grams of each sample were loaded into a vanadium sample holder inside a helium glove box. The total neutron scattering data for the doped samples were collected at temperatures of 15, 70, 125 and 300 K, with a collection time of four hours for each temperature point. The collected data were normalized using a vanadium standard measurement. The detector position and the wavelength of the neutron beam were calibrated with a Si measurement. Since HB-2A does not typically serve as a total scattering instrument, we processed the data further by performing a Fourier-transform using PDFgetN3~\cite{juhas2018pdfgetn3} with a \textit{Q}$\rm_{max}$ value of 10 Å$^{-1}$ to produce the PDF data. 

%%%%%%%%%%%%
% Begin Figure
%%%%%%%%%%%%
\begin{figure*}
    \centering
	\includegraphics[width=150mm]{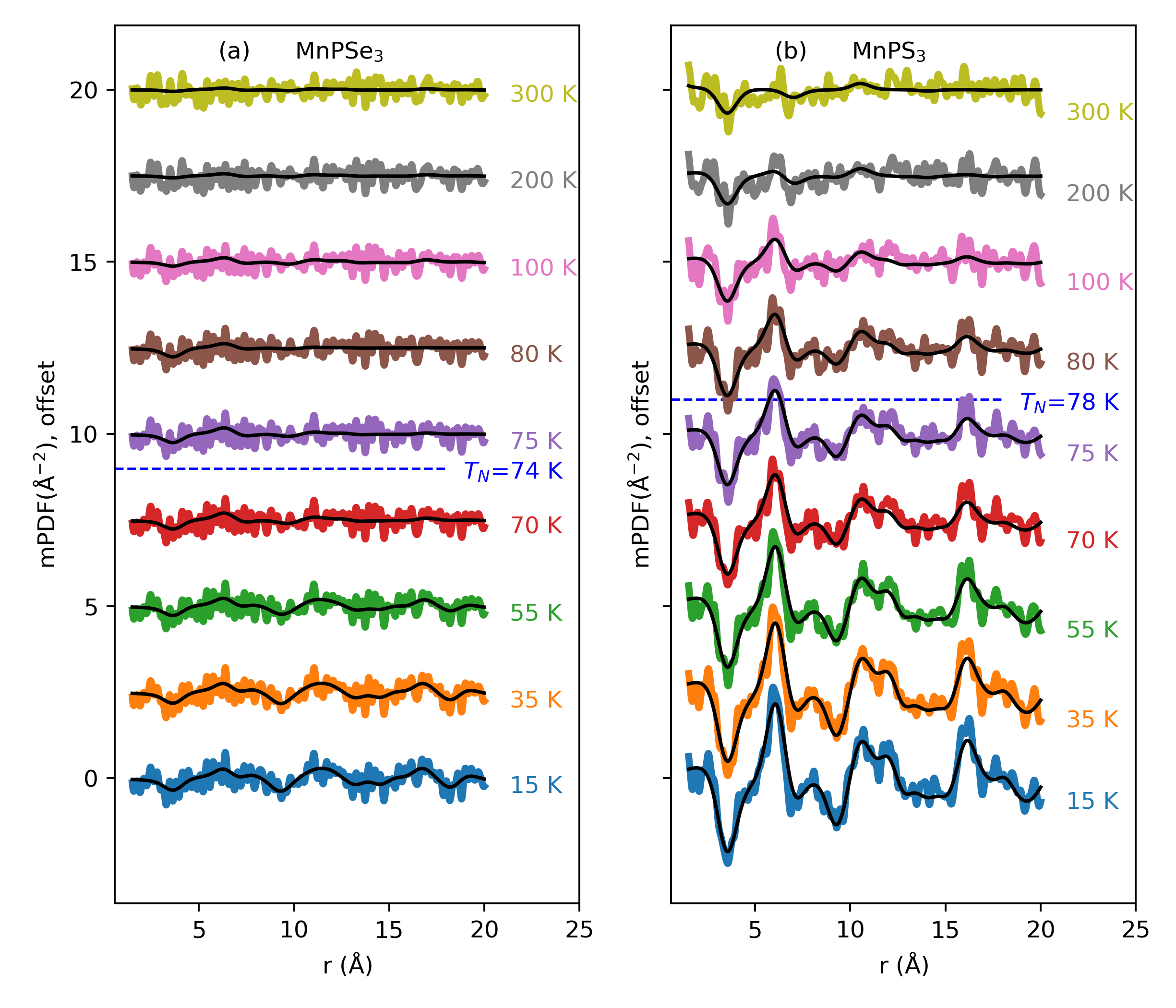}
	\caption{\label{fig:tseries_mpdf} Temperature series of the isolated magnetic PDF data and fits for \MPSe~(left) and \MPS~(right). The colored curves are the magnetic PDF data and the black curve represents the magnetic PDF fit (vertically offset for clarity). The long-range ordering transition temperature (\TN) is shown by the blue horizontal dashed line.
	}		
\end{figure*}
%%%%%%%%%%%%
% End Figure
%%%%%%%%%%%%

\subsection{Inelastic neutron scattering}

Inelastic neutron scattering (INS) measurements were conducted on MnPSSe$_{2}$, MnPS$_{1.5}$Se$_{1.5}$, and MnPS$_{2}$Se using the Time-of-Flight Direct Geometry Spectrometer Sequoia \cite{granroth2010sequoia} at the SNS, ORNL. An identical set-up was used compared to previous measurements on MnPSe$_{3}$ and MnPS$_{3}$ \cite{calder2021magnetic}. The samples were loaded into a cylindrical aluminum can with a diameter of 1 cm and INS spectra were measured using the three-sample changer within a closed-cycle refrigerator. Data were collected at incident energies of $E_{i}$ = 20 meV and $E_{i}$ = 8 meV, both in high-resolution mode. For $E_{i}$ = 20 meV, the Fermi chopper operated at a frequency of 240 Hz, with a $T_{0}$ chopper frequency of 60 Hz, resulting in an energy resolution of 0.48 meV at the elastic line. Similarly, for $E_{i}$ = 8 meV, the Fermi chopper frequency was 120 Hz, and the $T_{0}$ chopper operated at 30 Hz, providing an energy resolution of 0.18 meV at the elastic line. The energy resolution was determined through DGS resolution, confirmed by analyzing cuts of the elastic line within the data. Momentum resolution ($Q$) was derived by fitting the resolution of the Bragg peaks at the elastic line. Measurements were taken over six hours at temperatures of 200 K, 100 K, 70 K, and 15 K.

The collected neutron data underwent reduction using Mantid \cite{arnold2014mantid}. The data were adjusted for the Bose thermal population factor. Initial data reduction and viewing were carried out using the DAVE software \cite{azuah2009dave}. Modeling of the data  was performed using SpinW \cite{toth2015linear}.

\section{Results and discussion}	
\label{sec:results_discussion}
\subsection{Atomic and magnetic PDF analysis of \MPSe ~and \MPS}
\label{subsec:pdf_analysis}

We begin with neutron total scattering measurements on \MPSe ~and \MPS, with a focus on probing the spin ordering over a wide temperature range through the magnetic ordering transtions. This is achieved through atomic and magnetic PDF analysis of the neutron scattering data, which allows for the same analysis methodology to be applied in both the long range ordered phase and short range paramagnetic phase.
%We start with the unmixed samples, MnPSe$_{3}$ and MnPS$_{3}$,
Fig.~\ref{fig:mPDF_pure_15K}(a)-(b) shows the results at 15 K for MnPSe$_{3}$ and MnPS$_{3}$. The total scattering data include both atomic and magnetic scattering, which need to be disentangled. To isolate the magnetic data we employed an iterative fitting approach. Initially, we used the known atomic structure to perform the atomic PDF fit ~\cite{calder2021magnetic}. Then, we subtracted the calculated atomic PDF from the total PDF data to obtain the residual. This residual contains a magnetic signal which was used for the initial magnetic PDF refinement. Following this, we refined the atomic PDF again, but this time using the total PDF data minus the best-fit magnetic PDF signal for the atomic PDF input. The refined parameters in the first iteration served as starting points for the second iteration. We repeated this process for a third iteration, starting with the total PDF data minus the best-fit atomic PDF pattern from the second iteration, using the refined parameters from the second iteration. Finally, a third magnetic PDF fit was conducted. The detailed procedure of the iterative fitting approach of atomic and magnetic PDF is further discussed in the Supplementary section of Ref.\cite{baral2022real}. 

The analysis was performed on a temperature grid ranging from 15 K to 300 K and the atomic and magnetic PDF patterns were calculated over a real space fitting range of 1.5 - 20~\AA. For \MPSe~the low goodness of fit,~\Rw~=~0.0742, is evident from the small, featureless fit residual illustrated by the green curve at the bottom of the panel. When allowed to freely refine, the lowest \Rw~occurs when the spins are at an angle of 70$^{\circ}$ with the $c$-axis, however using the previously reported value of 90$^{\circ}$, where the spins are fixed to the $ab$-plane, produces a nearly identical fit \cite{calder2021magnetic}. Fig.~\ref{fig:mPDF_pure_15K}(b) shows the atomic and magnetic PDF results for \MPS, with~\Rw~ = ~0.139. The optimal magnetic fit occurs when the spins are aligned at an angle of 14.17 $\pm$ 2.09$^{\circ}$ with \textit{c}-axis, which is comparable to previous studies~\cite{ressouche2010magnetoelectric}.

We performed a series of atomic and magnetic PDF fits for various temperatures ranging from 15 to  300 K for both compositions. Fig.~\ref{fig:tseries_mpdf} shows the temperature series of the magnetic PDF fits for \MPSe~ and \MPS. For \MPSe~short range magnetic order persists above the N\'{e}el temperature (\TN = 74 K), up to 100 K and then decreases significantly beyond 100 K becoming almost unobservable at 200 K. Conversely in the case of \MPS, we observed a strong magnetic PDF signal well above the N\'{e}el temperature ( \TN = 78 K), with this local order magnetic signal persisting up to 300 K, which is the maximum temperature that the data were collected.

%%%%%%%%%%%%
% Begin Figure
%%%%%%%%%%%%
\begin{figure*}
    \includegraphics[width=165mm]{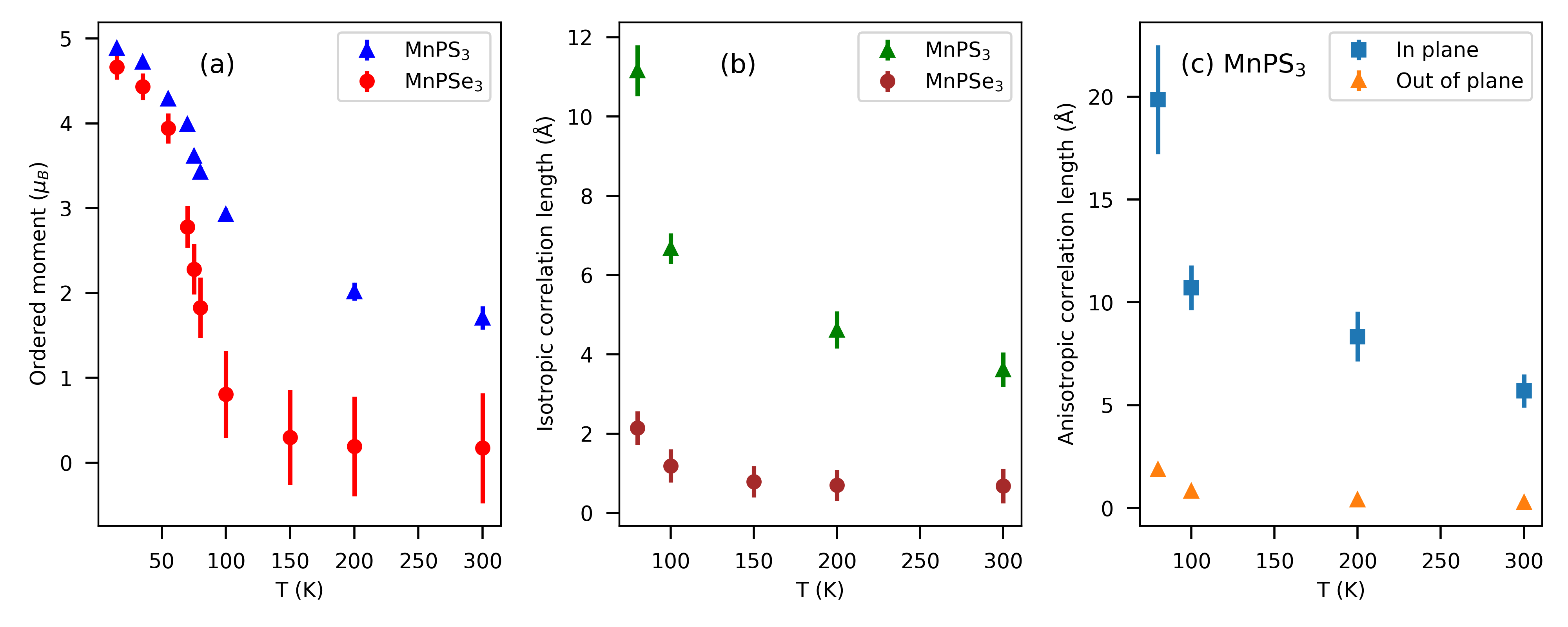}
    \caption{\label{fig:mT} 
     (a) Magnetic moments as a function of temperature for \MPSe~and \MPS, evaluated within a fitting range from 1.5 to 20 Å. The red and blue symbols represent the ordered moment for \MPSe~and \MPS, respectively.
    (b) Variation of the correlation length with temperature in the paramagnetic regime for MnPSe$_3$ and MnPS$_3$, based on isotropic fits in the range of 1.5 to 20 Å. (c) Best fit correlation lengths in the $ab$-plane (blue squares) and along the $c$-axis (orange triangles) as a function of temperature in a paramagnetic regime for \MPS using an anisotropic model.
    The error bars denote the standard deviation of the best-fit parameter values.}
\end{figure*}
%%%%%%%%%%%%
% End Figure
%%%%%%%%%%%%

%%%%%%%%%%%%%%%%%
% Begin Figure
%%%%%%%%%%%%%%%%%
\begin{figure*}
	\includegraphics[trim=0mm 0mm 0mm 0mm, clip,width=150mm]{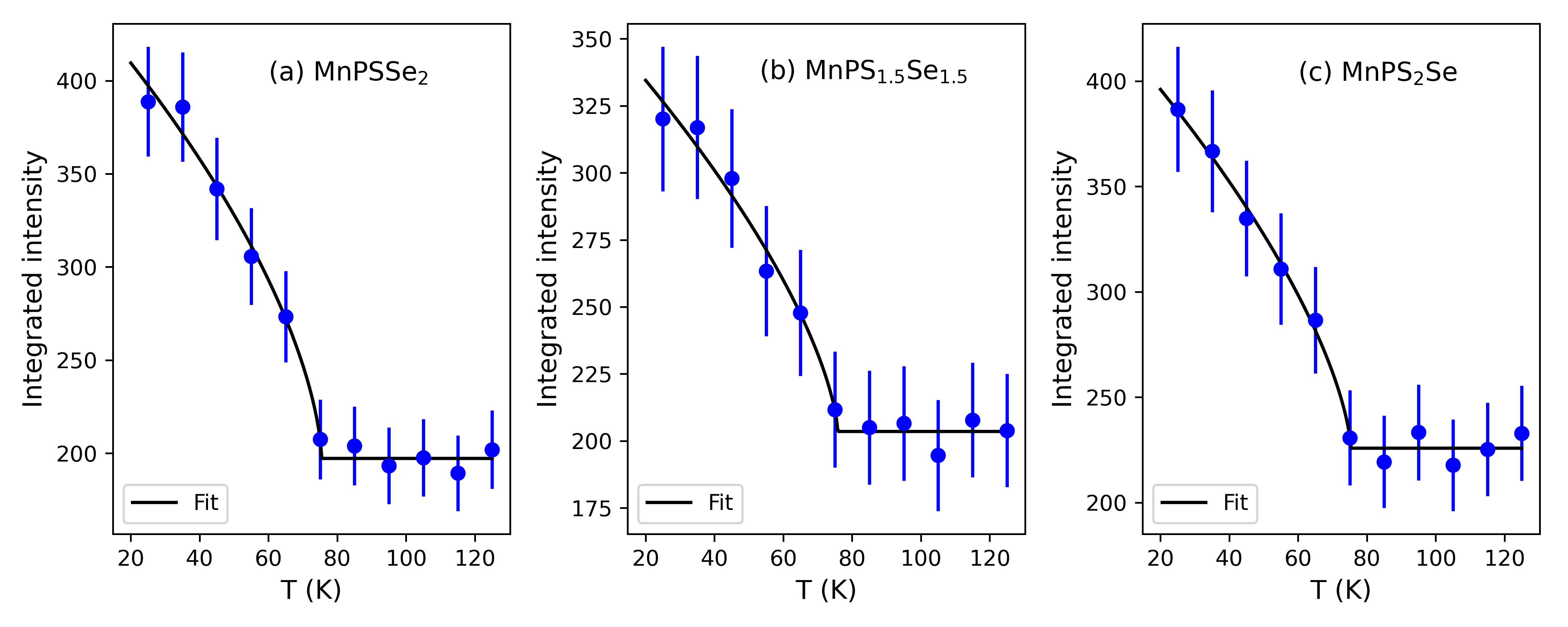}
	\caption{\label{fig:integrated_intensity}
     Power law fit to the integrated intensity of a magnetic Bragg peak in the range $Q$ = 2.68 to 2.73 $\rm A^{-1}$ for (a)  MnPSSe$_{2}$, (b) MnPS$_{1.5}$Se$_{1.5}$ and (c) MnPS$_{2}$Se.}
\end{figure*}
%%%%%%%%%%%%
% End Figure
%%%%%%%%%%%%

The mPDF analysis allows for a quantitative extraction of the magnetic correlation length and the ordered moment over a defined real space range.  The ordered moment in the fitting range 1.5 - 20~\AA~as a function of temperature for \MPSe~and \MPS~are shown in Fig.~\ref{fig:mT}(a). In the long range ordered regime the ordered moment begins to decrease exponentially as the sample temperature is increased towards the N\'{e}el temperature in both cases. With a further increase in temperature above the N\'{e}el temperature, the ordered moment further decreases. For \MPSe~the ordered moment approaches zero for temperatures of 150 K and above. For \MPS, however, while the ordered moment also decreases above the N\'{e}el temperature it attains a substantial moment of 1.7\muB~at 300 K, 1/3 of the total moment for Mn. The correlation length above the transition for both compositions, \MPS~and \MPSe, is illustrated in Fig.~\ref{fig:mT}(b). Below the transition temperature long-range order is established with an infinite correlation length. But above the transition with increasing temperature, the correlation length decreases for both compositions. \MPSe~has an uncorrelated moment at 150K, whereas the correlations persist in \MPS~to 300 K.

We conducted additional magnetic PDF fits for \MPS~above \TN~ over a fitting range 1.5 - 20~\AA~to capture the anisotropic magnetic correlations to further understand the persistence of local order. We implemented a model that distinguishes between correlation lengths in the $ab$-plane and along the $c$-axis. Our analysis reveals that the in-plane correlation ($\xi_{ab}$) length is approximately 15 times longer than out of plane correlation length ($\xi_{c}$) for all temperatures above \TN ( see Fig. \ref{fig:mT} (c)). This highlights that bulk \MPS~well approximates a 2D magnet. Undertaking the same anisotropic procedure for \MPSe~did not produce a clear distinction between the $ab$-plane and $c$-axis, within error. This indicates that the magnetism in \MPSe~is closer to three dimensional, despite the presence of a van der Waals gap between the layers.

Magnetic PDF measurements are not yet widely applied to 2D layered materials, but they can offer detailed insights. The results here show that \MPS~displays pronounced short-range magnetic correlations within the 2D layers well above the long range transition temperature, while this behavior is not found in \MPSe. This can be observed on a qualitative level from neutron diffraction data in reciprocal space, however mPDF analysis allows for a quantitative description in real space that can be applied to various ordering length scales both above and below the N\'{e}el temperature.

\subsection{Magnetic PDF analysis of MnPSSe$_{2}$, MnPS$_{1.5}$Se$_{1.5}$, and MnPS$_{2}$Se}

%%%%%%%%%%%%%%%%%
% Begin Figure
%%%%%%%%%%%%%%%%%
\begin{figure*}
	\centering
	\includegraphics[width=150mm]{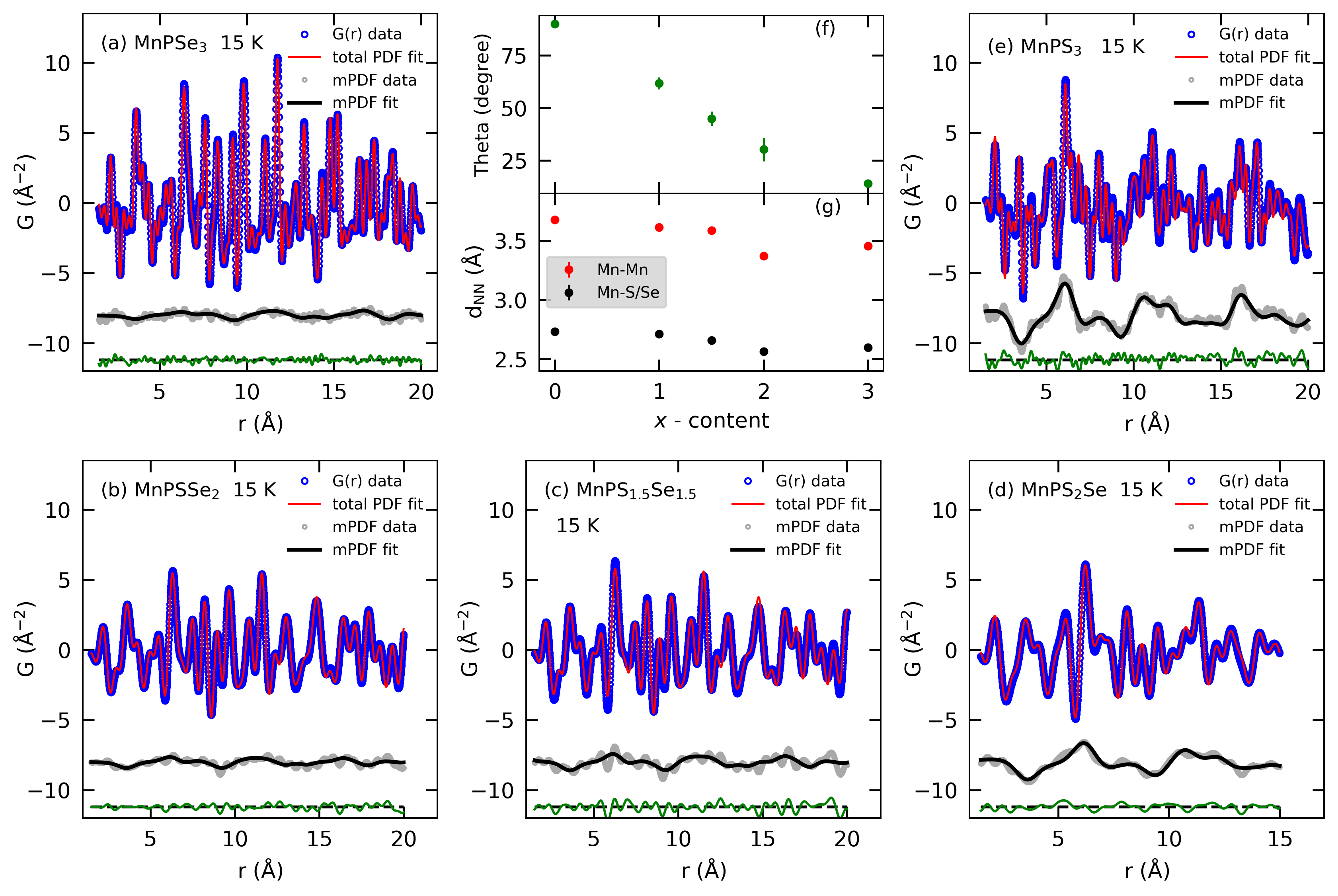}
	\caption{\label{fig:dopedlowTmPDFs} Total neutron and isolated magnetic PDF patterns of (a) \MPSe~(b) MnPSSe$_{2}$, (c) MnPS$_{1.5}$Se$_{1.5}$, (d) MnPS$_{2}$Se and (e) \MPS~ at 15~K. The blue, red, gray, black, and green symbols represent the observed PDF data, total PDF fit, isolated magnetic PDF data, magnetic PDF fit, and overall fit residual. (f)  Variation of spin angle (theta) with $x$-content for the series MnPS$_{x}$Se$_{3-x}$ at 15 K temperature. (g) Variation of nearest neighbor Mn-Mn and Mn-S/Se distances ($\text{d}_{\text{NN}}$) with $x$-content at 15 K for MnPS$_{x}$Se$_{3-x}$.}
\end{figure*}
%%%%%%%%%%%%
% End Figure
%%%%%%%%%%%%

% \begin{table}[tb]
%     \caption{Ordered magnetic moment and spin angle  with $c$-axis of  MnPS$_{x}$Se$_{3-x}$ ($x$ = 0, 1, 1.5, 2, 3) at 15 K. The spin-angle for MnPS$_3$ was fixed to 90 degrees, as described in the text.}
%     \label{tab:table}
%     \centering
%     \begin{tabular}{|l|r|r|} % Changed from { | c | c | c | }
%     \hline

%     Samples & ordered moment ($\mu_{B}$) & spin-angle (degree) \\ \hline
%     \MPS & 4.884 $\pm$ 0.036 & 14.17 $\pm$ 2.091 \\ 
%     MnPS$_{2}$Se & 4.313 $\pm$ 0.186 & 20.076 $\pm$ 5.382 \\ 
%     MnPS$_{1.5}$Se$_{1.5}$ & 4.370 $\pm$ 0.132 & 55.260 $\pm$ 3.534 \\
%     MnPSSe$_{2}$ & 4.470 $\pm$ 0.109 & 58.039 $\pm$ 2.529 \\
%     \MPSe & 4.666 $\pm$ 0.149 & 90 \\
%     \hline
%     \end{tabular}
% \end{table}

\begin{table}[tb]
    \caption{Ordered magnetic moment ($m$), spin-angle ($\theta$)  with $c$-axis, and N\'{e}el temperatures  of  MnPS$_{x}$Se$_{3-x}$ ($x$ = 0, 1, 1.5, 2, 3) at 15 K. The spin-angle for MnPS$_3$ was fixed to 90 degrees, as described in the text.}
    \label{tab:table}
    \centering
    \begin{tabular}{|l|c|c|c|} % Changed from { | c | c | c | c |}
    \hline

    Samples & $m$ ($\mu_{B}$) & $\theta$ (degree) & \TN~(K)\\ \hline
    \MPS & 4.884 $\pm$ 0.036 & 14.17 $\pm$ 2.091 & 78 Ref.~\cite{takano2004magnetic}\\ 
    MnPS$_{2}$Se & 4.313 $\pm$ 0.186 & 20.076 $\pm$ 5.382 & 75.31 $\pm$ 0.63\\ 
    MnPS$_{1.5}$Se$_{1.5}$ & 4.370 $\pm$ 0.132 & 55.260 $\pm$ 3.534 & 75.82 $\pm$ 1.35\\
    MnPSSe$_{2}$ & 4.470 $\pm$ 0.109 & 58.039 $\pm$ 2.529 & 75.42 $\pm$ 0.65 \\
    \MPSe & 4.666 $\pm$ 0.149 & 90 & 74 Ref.~\cite{wiedenmann1981neutron} \\
    \hline
    \end{tabular}
\end{table}

To gain further insight into the distinct behavior of \MPSe~and \MPS, we conducted atomic and magnetic PDF analysis on samples with controlled S/Se content of MnPSSe$_{2}$, MnPS$_{1.5}$Se$_{1.5}$, and MnPS$_{2}$Se to tune and investigate the behavior between the two materials. The same analysis procedure was applied to these samples as was used for \MPSe~and \MPS, however the measurements were conducted on different neutron scattering instruments. Data was collected at 15, 70, 125 and 300 K.
% , however it should be noted that $\rm Q_{max}=$ 10 $\rm \AA$$^{-1}$ was used for the doped samples in the reduction to produce PDF data due to the different neutron instruments used to collect the data compared to \MPSe~and \MPS.
The atomic and magnetic PDF calculations for MnPSSe$_{2}$, MnPS$_{1.5}$Se$_{1.5}$ were performed over the same fitting range of 1.5 - 20 \An~ for MnPSSe$_{2}$ and MnPS$_{1.5}$Se$_{1.5}$, while for MnPS$_{2}$Se, a different fitting range of 1.5 - 15 \An~ was selected. The variation of the fitting range on \S2~ was due to a notable degradation in the quality of the fit beyond 15 \An. There could be several factors for this, such as increased stacking faults or in-plane disorder from the site mixing that lead to this behavior. 

To obtain the magnetic ordering temperature of MnPSSe$_{2}$, MnPS$_{1.5}$Se$_{1.5}$, and MnPS$_{2}$Se we performed neutron diffraction measurements from 25 to 125 K and obtained the integrated intensity of a magnetic reflection, as shown in Fig.~\ref{fig:integrated_intensity}. Fitting this to a power law shows that the ordering temperature of MnPSSe$_{2}$, MnPS$_{1.5}$Se$_{1.5}$, and MnPS$_{2}$Se all occur within error around the same temperature of 75 K (see table~\ref{tab:table}), which is in between the ordering temperatures of 74 K for MnPSe$_3$ and 78 K for MnPS$_3$. 

The same iterative fitting approach was used to refine the atomic and magnetic PDF patterns as described in  subsection~\ref{subsec:pdf_analysis}. Fig.~\ref{fig:dopedlowTmPDFs} shows the total PDF and isolated magnetic PDF analysis patterns for all MnPS$_{x}$Se$_{3-x}$ ($x$ = 0, 1, 1.5, 2, 3) compositions at 15 K. 
%The atomic and magnetic PDF \MPSe, and \MPS~are also shown to allow for a comparison for all the compositions.
The best fit atomic and magnetic models for MnPSSe$_{2}$, MnPS$_{1.5}$Se$_{1.5}$ were obtained using the space group R$\bar{3}$ which is consistent with the space group of \MPSe. Similarly, the best atomic and magnetic fits for MnPS$_{2}$Se) were observed using the space group C${2}/m$, which corresponds to the space group for \MPS. Attempts to use alternative space groups did not produce improved fits for any compound. On comparing the mPDF patterns of all the compositions of $x$ = 0, 1, 1.5, 2, 3, it is observed that the real-space magnetic intensity slowly increases with $x$-content, suggesting an alteration of the magnetic order. Performing the mPDF fits to the isolated magnetic data reveals a gradual increase of the spin angle with alteration of the $x$ content in MnPS$_{x}$Se$_{3-x}$. Fig.~\ref{fig:dopedlowTmPDFs}(f) represents the variation of spin angle with $x$-content at 15 K. Table~\ref{tab:table} shows the variation of calculated order moment and the spin angle of MnPS$_{x}$Se$_{3-x}$ ($x$ = 0, 1, 1.5, 2, 3) at 15 K. As discussed in section~\ref{subsec:pdf_analysis} the spin angle for MnPSe$_3$ was indistinguishable between 70 degrees and the previously reported value of 90 degrees, as such 90 degrees is presented here.

The results of the mPDF analysis in Fig.~\ref{fig:dopedlowTmPDFs} show that a gradual change in the non-magnetic S/Se content is a parameter that can be used to directly control the magnetic spin direction. The underlying change in this process is to the crystal structure through chemical pressure as the different sized anions are altered. Fig.~\ref{fig:dopedlowTmPDFs}(g) shows the variation of nearest neighbor Mn-Mn and Mn-S/Se distances with $x$-content for MnPS$_{x}$Se$_{3-x}$ at 15 K. With progressive doping, both nearest neighbor Mn-Mn and Mn-S/Se distances decreases gradually with $x$-content. The decrease in the nearest neighbor Mn-Mn distance would be expected to lead to an increase in magnetic exchange interactions in the $ab$-plane. In addition, the change in Mn-S/Se distance will alter the crystal field environment and local anisotropy of the magnetic ion. 

\section{Inelastic Neutron Scattering}

%%%%%%%%%%%%%%%%%
% Begin Figure
%%%%%%%%%%%%%%%%%
%trim is left, bottom, right, top
\begin{figure*}[tb]
	\centering         
	\includegraphics[trim=0.0cm 0.8cm 0cm 0cm,clip=true, width=1.0\textwidth]{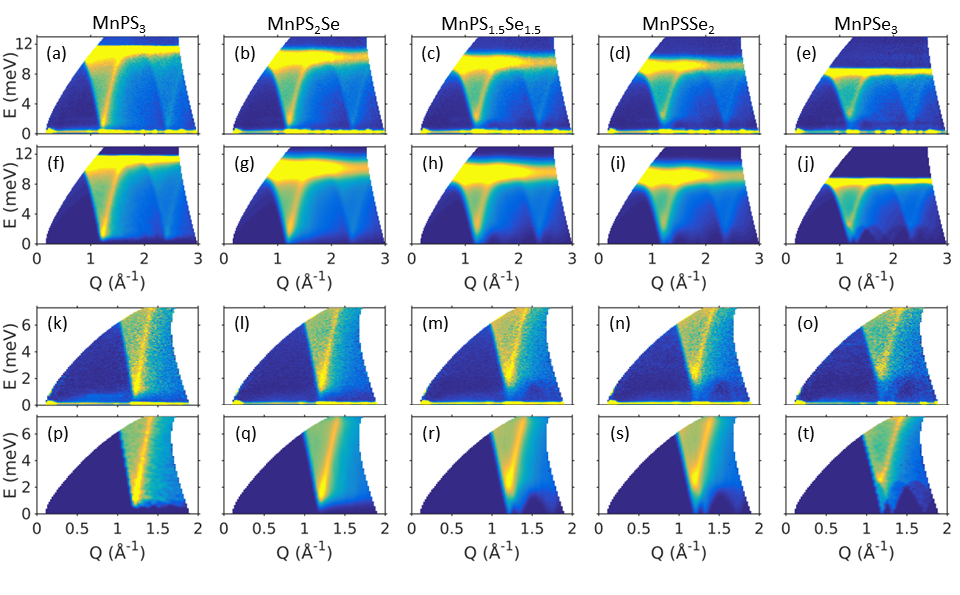}           
	\caption{\label{INS_data_model} Inelastic neutron scattering data and simulations for MnPS$_{x}$Se$_{3-x}$. Data collected with E$\rm _i$=20 meV for (a) MnPS$_3$, (b) MnPS$_2$Se, (c)   MnPS$_{1.5}$Se$_{1.5}$, (d)  MnPSSe$_{2}$ and (e)  MnPSe$_{3}$. (f)-(j) Best fit models using linear spin wave theory. Data collected with E$\rm _i$=8 meV for (k) MnPS$_3$, (l) MnPS$_2$Se, (m)   MnPS$_{1.5}$Se$_{1.5}$, (n)  MnPSSe$_{2}$ and (o)  MnPSe$_{3}$. (p)-(t) Best fit models using linear spin wave theory.}
\end{figure*} 
%%%%%%%%%%%%
% End Figure
%%%%%%%%%%%%

To investigate the influence of altering the non-magnetic S and Se content on the magnetic exchange interactions in the series MnPS$_{x}$Se$_{3-x}$ further we performed INS measurements on the powder samples. Measurements of the undoped compounds MnPSe$_3$ and MnPS$_3$ have been previously performed to extract the exchange interactions \cite{calder2021magnetic,wildes1998spin}, these values are shown in Table.~\ref{Table_INS}. Comparing the values shows that MnPS$_3$ has stronger in-plane interactions, however the interlayer interactions are an order of magnitude smaller than MnPSe$_3$. This places the bulk MnPS$_3$ compound closer to the ideal 2D limit. This behavior was observed in the mPDF analysis reported above. In addition MnPS$_3$ has spin anisotropy, as evidenced by a measurable spin-gap in INS \cite{wildes1998spin}, which was modeled with single-ion anisotropy. No such spin-gap was observed in MnPSe$_3$. Given these difference in the exchange interactions, it is somewhat surprising that MnPS$_3$ and MnPSe$_3$ order within a few degrees K of each other. The stronger (weaker) $ab$-plane interactions and weaker (stronger) interlayer interactions for MnPS$_3$ (MnPSe$_3$) appear to lead to a balance that results in similar ordering temperatures. How these interactions change in-between the two parent compounds is the focus of the  INS measurements on MnPS$_{x}$Se$_{3-x}$ ($x=1, 1.5, 2$).   

The INS data for the full series MnPS$_3$, MnPS$_2$Se,  MnPS$_{1.5}$Se$_{1.5}$,  MnPSSe$_{2}$,  MnPSe$_{3}$ are shown in Fig.~\ref{INS_data_model}. An evolution is observed in the magnetic excitation spectra as the non-magnetic S and Se ion contents are altered. MnPS$_3$ has the widest energy bandwidth, indicating stronger magnetic exchange interactions, and this gradually reduces as S is replaced with Se. In addition, there is an evolution in the low energy scattering, with the reported spin-gap in MnPS$_3$ reducing and then becoming unobservable. This is replaced with low energy scattering revealing appreciable interlayer exchange interactions, as observed most strongly in MnPSe$_{3}$. 

To extract quantitative information on MnPS$_2$Se,  MnPS$_{1.5}$Se$_{1.5}$ and  MnPSSe$_{2}$ we follow a similar modeling process as that performed in Ref.~\cite{calder2021magnetic}. We use linear spin wave theory with a model based on a 2D Hamiltonian on a honeycomb lattice,

\begin{eqnarray*}
	\mathcal{H}=  \sum_{ i \ne j}J_{ij} \mathbf{S}_i\cdot\mathbf{S}_j =\sum_{\langle ij \rangle}2J_{ij} \mathbf{S}_i\cdot\mathbf{S}_j
\end{eqnarray*}

that included $J_{1ab}$, $J_{2ab}$ and $J_{3ab}$ intralayer exchange interactions, as well as the full spin (S=5/2). Note the exchange interactions have been reported to be consistent with parameters in the literature and allow comparisons between MnPSe$_3$ and MnPS$_3$ \cite{wildes1998spin,PhysRevB.91.235425,yang2020electronic,PhysRevB.94.184428}. Alternative notation related by $J$/2 is used for MnPS$_3$ in Ref.~\cite{Vaclavkova_2020}.

The top of the band for MnPS$_3$ and MnPSe$_3$ are sharp and resolution limited for the E$\rm _i$=20 meV measurements. There is some broadening for the MnPS$_2$Se,  MnPS$_{1.5}$Se$_{1.5}$ and  MnPSSe$_{2}$. This can be attributed to a level of structural disorder introduced in the S and Se mixed compounds which would damping the excitations, as often observed in doped samples. This was accounted for in the modeling by broadening the energy resolution for the $x=1, 1.5, 2$ analysis. A series of 1D cuts were taken at constant Q and constant E during the fitting process. To restrict the number of variables to allow for the extraction of a minimal model the $J_{1ab}$, $J_{2ab}$ and $J_{3ab}$ interactions for MnPSe$_3$ and MnPS$_3$ where used as starting values. A scale factor was applied to model the top of the band of magnetic excitations. These values were then fixed and the interlayer interaction ($J_c$) and anisotropy in the form of SIA were altered in the subsequent modeling. The determined values are shown in Table.~\ref{Table_INS} and the comparison of experimental data and model are shown in Fig.~\ref{INS_data_model}.

The data and modeling shows a consistent trend of reducing the in-plane interactions and increasing the interlayer interactions as the S content is diluted with Se. This trend in exchange interactions and magnetic excitation spectra gradually altering agrees with the gradual switching of spins from in-plane (MnPSe$_3$) to almost fully along the c-axis in (MnPS$_3$) with altering S/Se content as found in the mPDF analysis.

\begin{table}[tb]
	\caption{\label{Table_INS}Exchange interactions from modeling the INS data with linear spin wave theory. All values are in meV.}
	\begin{tabular}{ |l | r | r | r | r | r |}
		\hline 
		           &           &              &                        &              &              \\ 
		           &  MnPS$_3$ &  MnPS$_2$Se  & MnPS$_{1.5}$Se$_{1.5}$ & MnPSSe$_{2}$ & MnPSe$_{3}$  \\ 
	               & Ref.~\cite{wildes1998spin} &              &                        &              & Ref.~\cite{calder2021magnetic}    \\ \hline
		$J_{1ab}$  &  0.77     &   0.716      &   0.529                & 0.501        & 0.45         \\ 
		$J_{2ab}$  &  0.07     &   0.065      &   0.035                & 0.033        & 0.03         \\ 
		$J_{3ab}$  &  0.18     &   0.167      &   0.218                & 0.207        & 0.19         \\ 
		$J_c$      &  -0.0038  &   --         &   0.019(4)             & 0.025(3)     & 0.031        \\ 
	 	SIA        &  0.0086   &   0.0045(15) &   --                   & --           & --           \\ 

		\hline
	\end{tabular}
\end{table}

\section{Conclusion}

In summary, the magnetic behavior of a series of 2D layered van der Waals materials MnPS$_{x}$Se$_{3-x}$ ($x$ = 0, 1, 1.5, 2, 3)  were investigated through atomic and magnetic PDF analysis and inelastic neutron scattering. The magnetic ordering temperature was found to remain  within a few Kelvin (74 K - 78 K) for the whole series, however the spin direction, correlation length and exchange interactions evolve over a wide range as $x$ is tuned.

The mPDF signal was isolated from the total scattering data to follow the magnetic behavior over a wide range of temperature and $x$ composition. The mPDF results revealed short-range antiferromagnetic correlations in the paramagnetic regime. \MPS~was found to have a significantly longer correlation length  compared to \MPSe, with short range order in the 2D layers existing up to room temperature. A progressive increase in Se-doping leads to a notable shift in spin orientation from out-of-plane to in-plane. This is found to occur gradually with $x$ concentration. 

The exchange interactions were parameterized by a minimal model antiferromagnetic Heisenberg Hamiltonian on the honeycomb lattice. These showed a gradual change in the in-plane interactions, which were largest for \MPS~ and decreased with increasing $x$. Conversely the interlayer interactions are largest for \MPSe~and increase with $x$. 

The change in the magnetism in MnPS$_{x}$Se$_{3-x}$ ($x$ = 0, 1, 1.5, 2, 3) was induced by altering the non-magnetic S/Se ions which alters the chemical pressure, therefore structural control is central to the magnetic behavior. This highlights the ability to design and tune the magnetism in the 2D layers of MnPS$_{x}$Se$_{3-x}$ by not only controlling the $x$, but also through other perturbations including strain and pressure. More generally this investigation shows how the application of real space analysis of neutron data with mPDF can provide insights into 2D layered van der Waals materials where understanding local and long range ordering over a wide range of temperature of structural composition is often of interest.

\section*{acknowledgments}
The authors thank Benjamin Frandsen for the useful discussion. This research used resources at the High Flux Isotope Reactor and Spallation Neutron Source, a DOE Office of Science User Facility operated by the Oak Ridge National Laboratory. D.M. acknowledges support from the Gordon and Betty Moore Foundation’s EPiQS Initiative, Grant GBMF9069. This manuscript has been authored by UT-Battelle, LLC under Contract No. DE-AC05-00OR22725 with the U.S. Department of Energy. The United States Government retains and the publisher, by accepting the article for publication, acknowledges that the United States Government retains a non-exclusive, paidup, irrevocable, world-wide license to publish or reproduce the published form of this manuscript, or allow others to do so, for United States Government purposes. The Department of Energy will provide public access to these results of federally sponsored research in accordance with the DOE Public Access Plan(http://energy.gov/downloads/doepublic-access-plan).

%\textbf{Author Contributions}

%\textbf{Declaration of Interests}
%The authors declare no competing interests.


\end{document}